\documentclass[aps,prb,twocolumn]{revtex4} 

\usepackage{graphicx}
\usepackage{dcolumn}
\usepackage{bm}

\newcommand{\comment}[1]{}

\begin{document}
\renewcommand{\theequation}{\arabic{section}.\arabic{equation}}

\title{Attraction Between Electron Pairs
in High Temperature Superconductors}


\author{Phil Attard}
\affiliation{{\tt phil.attard1@gmail.com}
\\ 4 Mar., 2022} 


\begin{abstract}
It is proposed that in high temperature superconductors Cooper pairs
form and condense due to the monotonic-oscillatory transition
in the  pair potential of mean force,
which occurs quite generally at high coupling in charge systems.
It is shown that the predicted transition temperatures
are broadly in line with measured superconducting transition temperatures
for reasonable values of the total electron density
and the residual dielectric permittivity
arising from the immobile electrons.
The predicted transition is independent of the isotopic masses of the solid.
Consequent design principles
for high temperature superconductors are discussed.
\end{abstract}


\maketitle

%
\section{Introduction}
\setcounter{equation}{0} \setcounter{subsubsection}{0}
%

The Bardeen-Cooper-Schrieffer (BCS)\cite{BCS57}
theory of superconductivity is based on
electron pairs forming zero-momentum, zero-spin bosons
bound together by an attractive interaction.
For normal low temperature superconductors
the accepted mechanism for the attraction
is the dynamic interaction of the electron pair with phonons,
which are the quantized vibrations of the solid lattice.
The dependence of the transition temperature
on the isotopic masses of the solid\cite{Maxwell50,Reynolds50}
confirms BCS theory.

In the late 1980's high temperature superconductors
with transition temperatures above 30\,K were discovered.
\cite{Bednorz86,Wu87}
These are independent of the isotopic masses,
which rules out the phonon exchange mechanism.
New materials
with ever higher superconducting transition temperatures
have been discovered in the ensuing years,
but despite many proposals,
\cite{Anderson87,Bickers87,Inui88,Gros88,Kotliar88,Mann11,Monthoux91}
no consensus has emerged for the nature of the attractive potential
that forms Cooper pairs in these materials.

In this paper I propose that the electron attraction
responsible for Cooper pairs in high temperature superconductors
is due to the oscillatory pair static correlation function
that occurs at high coupling.
This has long been established
for like-charged particles
in the one component plasma
and in primitive model electrolytes.
\cite{Brush66,Stillinger68,Fisher69,Outhwaite78,Stell76,Parrinello79,%
Attard93,Ennis95}
Here the high-temperature superconductor
is modeled as a one-component plasma \emph{in media},
with the relatively few electrons in the Fermi foam
comprising the fluid charges,
and the fixed nuclei and majority immobile electrons in the Fermi sea
forming the neutralizing background and static relative permittivity
(dielectric constant).
The  predicted temperatures for the monotonic-oscillatory transition
in this model encompass the measured transition temperatures
for high temperature superconductors
for physically reasonable values of the total electron density
and static relative permittivity.

%
\section{Model and Analysis}
\setcounter{equation}{0} \setcounter{subsubsection}{0}
%

\comment{ 
As mentioned,
I model the solid conductor
as a one component plasma \emph{in media},
with the mobile electrons belonging to the Fermi foam,
and the fixed atoms
and immobile electrons,
which belong to the Fermi sea,
giving rise to the neutralizing background
and the static relative permittivity.
I first give an analytic expression
for the location of the monotonic-oscillatory
transition in the restricted primitive model electrolyte,
which I show has the same functional form
as the numerical result for the one component plasma.
This allows me to estimate the width of the Fermi foam
and hence the density of mobile electrons,
which is required to locate the transition temperature.
The final result requires to physical parameters:
the residual static relative permittivity,
and the total density of electrons in the solid.
}  

The solid conductor is modeled as a one component plasma
(mobile electrons in a uniform counter-charge background),
together with a finite relative permittivity,
$  \epsilon_\mathrm{r} = {\cal O}(10^2) $,
that results from the remaining immobile but polarizable electrons
(i.e.\ those deep in the Fermi sea).
The mobile electrons at the Fermi foam
have number density $\rho_\mathrm{F}(T)$ derived below.
Arguments concerning this model are canvassed in the conclusion.

I begin with the restricted primitive model electrolyte
for three reasons:
First, there are a wealth of analytic, numeric, and experimental results
known for electrolytes.
Second, it shows the generality of the monotonic-oscillatory transition
in charge systems.
And third,
it gives a specific value for the width of the accessible energy states,
which is required to determine the electron density of the Fermi foam.

In the restricted primitive model electrolyte
(ions of equal hard sphere diameter),
the pair distribution function
undergoes an oscillatory transition  when\cite{Attard93}
\begin{equation} \label{Eq:kDd=2}
\kappa_\mathrm{D} d \ge \sqrt{2} ,
\end{equation}
where $d$ is the hard core diameter of the ions.
The inverse Debye screening length for the binary symmetric electrolyte is
$\kappa_\mathrm{D}
= \sqrt{ (4\pi\beta/\epsilon) 2 \rho_\mathrm{F}  q^2 }$,
where $q$ is the ionic charge (in this case the electron charge),
and $\rho_\mathrm{F}$ is the number density of each type of ion.
Here $\beta = 1/k_\mathrm{B}T$ is the inverse temperature,
and $\epsilon = 4\pi \epsilon_0 \epsilon_\mathrm{r}$ is the
total permittivity of the medium,
$\epsilon_0$ being the permittivity of free space (SI units).
This result
is based on the Debye-H\"uckel form for
the pair distribution function
combined with the exact Stillinger-Lovett second moment condition.
More accurate analytic and numeric approximations exist,\cite{Attard93}
but this is sufficient for the present purposes.

To make the connection
with the one component plasma,
which does not impose a hard core diameter,
the distance of closest approach of the electrons can be set
as the point at which the Coulomb potential  \emph{in media} reaches
several times the thermal energy, $u(d) = \alpha k_\mathrm{B}T$,
or $d = \beta e^2/\epsilon \alpha$.
With these the oscillatory transition in the symmetric electrolyte
occurs when
\begin{equation}
2 \le
\frac{4\pi\beta 2\rho_\mathrm{F} e^2 }{\epsilon}
\frac{ \beta^2 e^4}{\epsilon^2 \alpha^2}
= \frac{6}{ \alpha^2 } \Gamma^3 .
\end{equation}
Here the plasma coupling parameter with finite relative permittivity
is $\Gamma \equiv \beta e^2/[\epsilon (3/4\pi \rho_\mathrm{F})^{1/3}]$.

As mentioned, the parameter $\alpha$ is the multiple of the thermal energy
which bounds the accessible states.
Choosing $\alpha = \surd 24 \approx 4.9$, the transition criterion becomes
\begin{equation} \label{Eq:G=2}
\Gamma \ge 2 .
\end{equation}
With this value of $\alpha$,
the value of the coupling constant at the transition
given for the restricted primitive model electrolyte
agrees with that found by Monte Carlo simulations
of the one component plasma.\cite{Brush66}

I now estimate the density of the electrons in the Fermi foam
modeling them as a  non-interacting ideal gas,
also known as the free electron model.
The Fermi momentum and the Fermi energy for ideal fermions
are\cite{Pathria72}
\begin{equation}
p_\mathrm{F} =
2\pi\hbar \left(\frac{3\rho}{8\pi}\right)^{1/3},
\mbox{ and }
\epsilon_\mathrm{F} =
\frac{(2\pi\hbar)^2 }{2 m} \left(\frac{3\rho}{8\pi}\right)^{2/3} ,
\end{equation}
where $\rho = N/V$ is the total electron number density.
The thermal wavelength is $\Lambda = \sqrt{2\pi \beta \hbar^2 /m}$,
and $\beta \epsilon_\mathrm{F} =2\pi ( 3\rho \Lambda^3 /8\pi)^{2/3}/2$,
which is much larger than unity.

With momentum state spacing being $\Delta_p = 2\pi \hbar /L$,
\cite{Messiah61,Merzbacher70}
where the volume is $V=L^3$,
the number in the Fermi foam is
\begin{eqnarray}
N_\mathrm{F}
& = &
2 \Delta_p^{-3}
\int_{\epsilon_\mathrm{F} - \alpha/\beta}^{\epsilon_\mathrm{F} + \alpha/\beta}
\mathrm{d} \epsilon\, 4\pi  m \sqrt{2m\epsilon}
\frac{e^{-\beta (\epsilon - \epsilon_\mathrm{F})}
}{ 1 + e^{-\beta (\epsilon - \epsilon_\mathrm{F})} }
\nonumber \\ & \approx &
4 \alpha V \Lambda^{-3}
( 3\rho \Lambda^3 /8\pi)^{1/3} .
\end{eqnarray}
An expansion to leading order for large  $\beta \epsilon_\mathrm{F} $
has been made to obtain the final equality.
That is
\begin{equation} \label{Eq:rhoF}
\rho_\mathrm{F} \Lambda^3
 =
4 \alpha \left(  \frac{3\rho \Lambda^3}{8\pi} \right)^{1/3} .
\end{equation}
The total excitable electron density, $\rho_\mathrm{F}$,
is significantly less than the total electron density, $\rho$.
It is proportional to the number of accessible energy states
at the Fermi energy,
which is fixed by equating the results of the restricted primitive model
to those of the one component plasma,
$\alpha = \surd 24 $

The idea that only the electrons at the Fermi surface contribute
to screening also underlies the Thomas-Fermi model of the electron gas.
\cite{Kittel76}
This idea is taken a little further here by modeling the remaining
immobile electrons as being polarisable
and contributing to the residual dielectric constant.


%
\section{Numerical Results}
\setcounter{equation}{0} \setcounter{subsubsection}{0}
%

The predicted oscillatory-monotonic transition temperature
is now explored for a range of the two free parameters:
the total electron density $\rho$ and
the residual static relative permittivity $\epsilon_\mathrm{r}$.
For these a guide is provided by values for ceramic materials.
The total electron density of zirconia ZrO$_2$ is
$\rho = 1.65 \times 10^{30}$\,m$^{-3}$.
The relative permittivity of typical ceramic insulators
is on the order of $\epsilon_\mathrm{r} =$ $10^1$--$ 10^2$. 
\cite{ceram}

\begin{figure}[t!]
\centerline{ \resizebox{8cm}{!}{ \includegraphics*{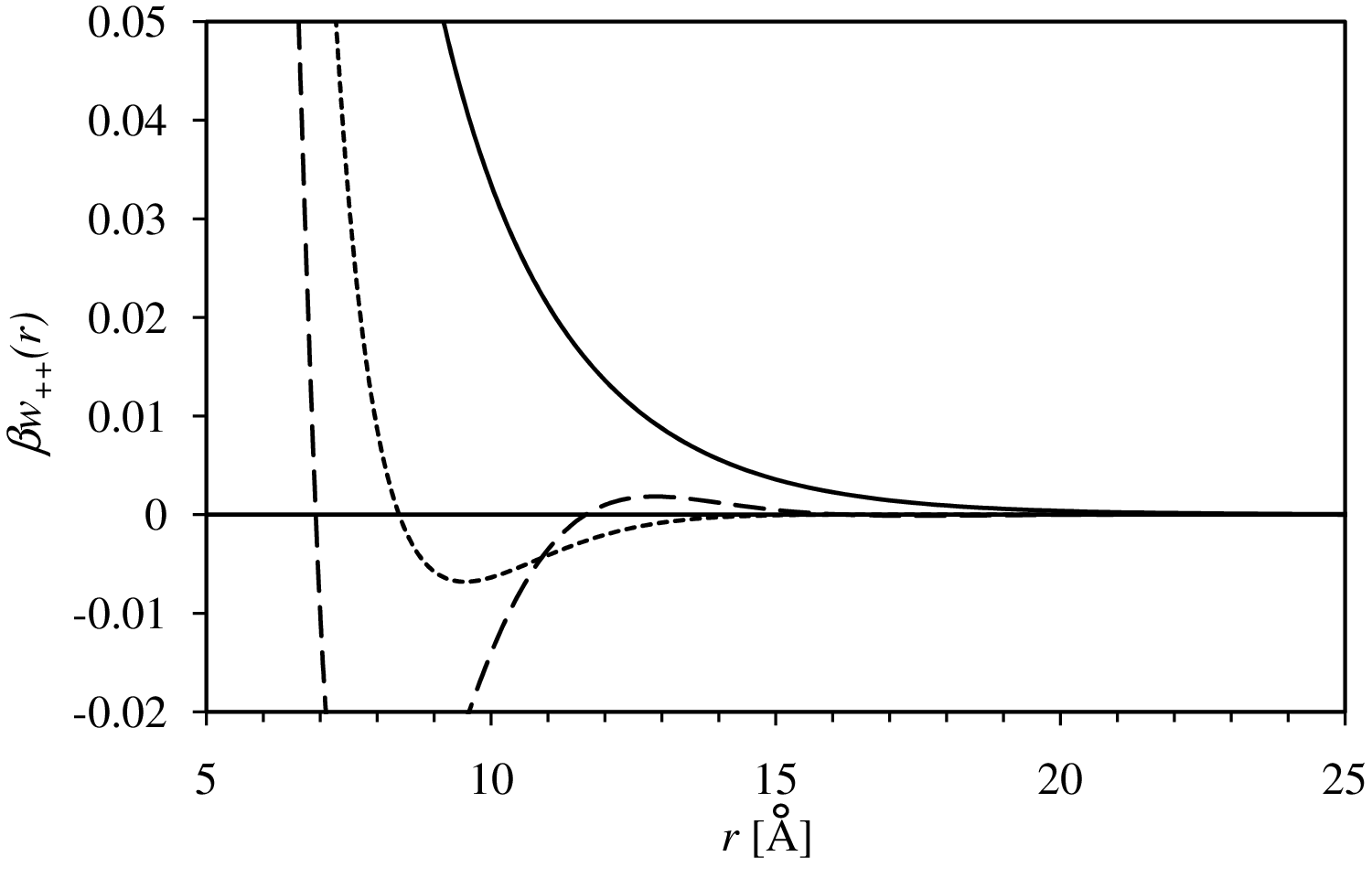} } }
\caption{\label{Fig:pmf}
Coion pair potential of mean force
as a function of separation in
the symmetric binary monovalent electrolyte
($d=3.41$\,\AA, $\epsilon_\mathrm{r} = 100$,  $T=100$\,K,
hypernetted chain approximation).
The solid curve is 0.5\,M,
($\kappa_\mathrm{D}^2d^2 = 1.5$, $\Gamma = 1.8$), 
the short-dash curve is  1.0\,M,
($\kappa_\mathrm{D}^2d^2 = 2.9 $, $\Gamma = 2.3$), 
the long-dash curve is for 2.0\,M
($\kappa_\mathrm{D}^2d^2= 5.9 $, $\Gamma = 2.9$). 
The solid line is an eye guide.
}
\end{figure}

Figure~\ref{Fig:pmf} shows the coion pair potential of mean force
above and below the oscillatory transition
close to contact.
The results were obtained with the hypernetted chain approximation
for the restricted primitive model binary electrolyte.
\cite{Attard93}
The qualitative difference that the transition makes is apparent.
Increasing coupling corresponds to decreasing temperature
at constant concentration,
or increasing concentration  at constant temperature,
which is the case shown in the figure.

At the lowest concentration shown  0.5\,M,
the potential of mean force is monotonic repulsive and exponentially decaying.
In this case the coupling is below the transition value
predicted for the restricted primitive model
and also below that predicted for the one component plasma.
At the intermediate concentration in the figure, 1.0\,M,
a shallow primary minimum appears with width on the order of $10^3$\,\AA.
At the highest concentration shown, 2.0\,M,
the primary minimum has become relatively deep and narrow,
and the potential of mean force is clearly oscillatory
with a noticeable barrier to the primary minimum.
For the present parameters,
the location of the transition predicted
by the Debye-H\"uckel--Stillinger-Lovett approximation
for the primitive model electrolyte,
Eq.~(\ref{Eq:kDd=2}),
is more or less equal to that observed in the one component plasma,
Eq.~(\ref{Eq:G=2}).

The molecular interpretation of the attraction
is that it arises at high coupling from over-charging by counterions
(or the background charge)
combined with packing constraints.\cite{Attard93}
As mentioned, it also occurs in the one component plasma.\cite{Brush66}

\comment{ 
One might crudely approximate the potential minimum
at the lowest temperature shown in Fig.~\ref{Fig:pmf}
as a square well of width $a=5$\,\AA\
and depth $v_0 =0.3\,k_\mathrm{B}T = 2 \times 10^{-22}$\,J.
Estimating the kinetic energy as
$\hbar^2/ma^2 = 5\times 10^{-20}$\,J
suggests that the electron pair is not actually bound
in a quantum mechanical sense.
Of course this estimate is rather crude,
and one can question the quantitative applicability
of the primitive model calculations.
More important conceptually is the point
that quantum mechanics is not the appropriate theory for condensed matter.
As is shown in the accompanying paper,
quantum statistical mechanics should be applied instead,
in which theory Cooper pairs do not require bound states.
} 

\begin{figure}[t!]
\centerline{ \resizebox{8cm}{!}{ \includegraphics*{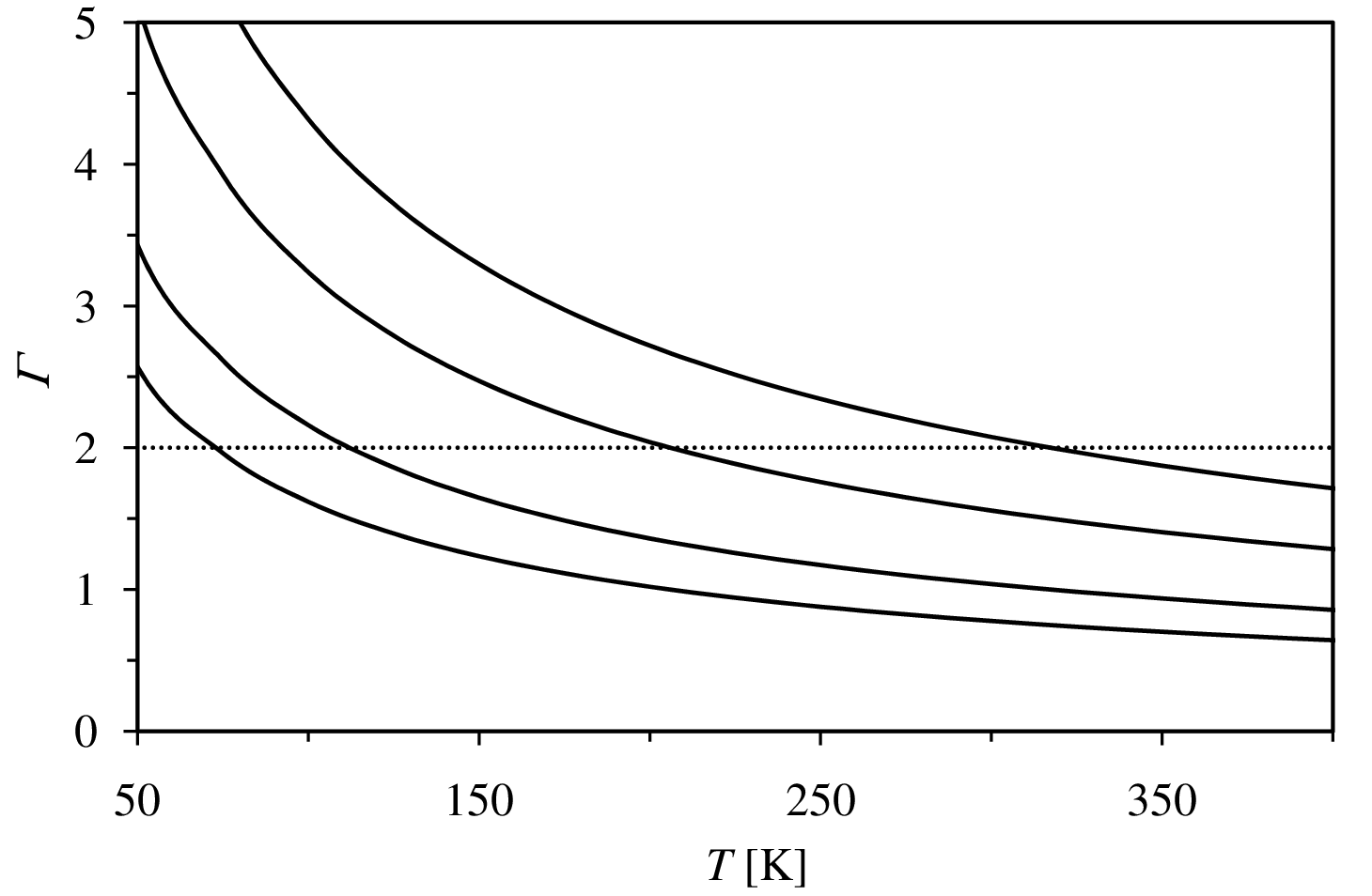} } }
\caption{\label{Fig:GvsT}
Plasma coupling parameter \emph{in media}
using electron density $\rho_\mathrm{F}(T)$, Eq.~(\ref{Eq:rhoF}),
and, from bottom to top, relative permittivity
$\epsilon_\mathrm{r} =$ 200, 150, 100, and 75.
The dotted line marks the transition to oscillatory behavior.
In all cases $\alpha=\surd 24$ and $\rho = 10^{30}$\,m$^{-3}$.
}
\end{figure}

Figure~\ref{Fig:GvsT}
shows the value of the plasma coupling parameter
as a function of the relative permittivity
for fixed total electron density.
Values $\Gamma > 2$ mark oscillatory pair correlation functions.
One sees that this occurs for low temperatures.
For the lowest relative  permittivities shown,
$\epsilon_\mathrm{r} =$ 100 and 75,
the transition temperatures are greater than
those measured for high temperature superconductors.
\cite{Bednorz86,Wu87}

\begin{figure}[t!]
\centerline{ \resizebox{8cm}{!}{ \includegraphics*{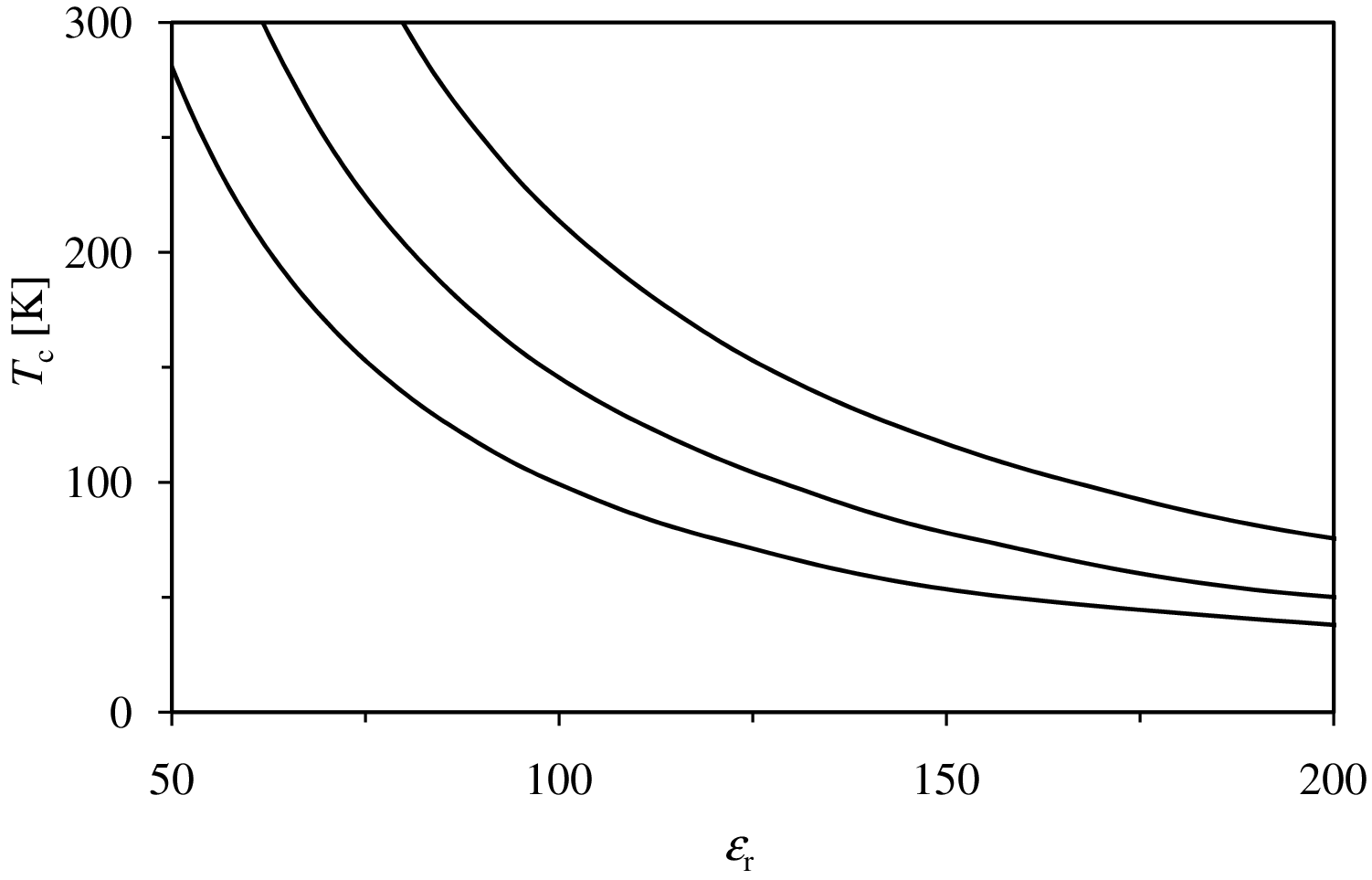} } }
\caption{\label{Fig:Tc}
Transition temperature as a function of the relative permittivity.
From left to right the curves are for a total electron density of
$\rho =$ 0.1, 1, and 10 $\times 10^{30}$\,m$^{-3}$.
}
\end{figure}

Figure~\ref{Fig:Tc} shows the transition temperature
as a function of relative permittivity
for several values of the electron density.
The transition temperature increases with decreasing relative permittivity
and with increasing electron density.
It is more sensitive to changes in the permittivity
than the electron density.
The range of calculated transition temperatures
encompasses those measured for high temperature superconductors.
From the figure one can conclude
that a high superconducting transition temperature requires
both a high electron density
and a low static relative permittivity.
Since the latter can be expected to increase with increasing electron density,
there is an obvious competition between these two requirements.
This undoubtedly restricts the structure and composition
of suitable candidate materials for high temperature superconductivity.

%
\section{Conclusion}
\setcounter{equation}{0} \setcounter{subsubsection}{0}
%

Perhaps the most controversial aspect of the proposed model
is the invocation of a finite relative permittivity for the solid.
The perceived wisdom  is that a conductor has infinite permittivity.
Also, the one component plasma is usually modeled \emph{in vacuo},
$\epsilon_\mathrm{r} = 1$.
In response to these anticipated objections I make the following points.
First, one has to distinguish between an experimental challenge
and a fundamental limitation of nature.
There is no doubt that a macroscopic measurement of the dielectric constant
of a conductor yields an infinite dielectric constant
in the zero frequency limit.
In my opinion, all this says
is that the conductivity dominates the measurement;
it does not say that the residual relative permittivity  is
either unity or infinity.
Second,
fixed atoms surrounded by the electrons deep in the Fermi sea,
which include the immobile inner shell electrons,
remain polarizable,
and therefore they must contribute to a finite relative permittivity.
Third,
if the relative permittivity were truly infinite at the molecular level,
then the immobile electrons could not interact via the Coulomb repulsion,
and they would be utterly transparent to each other,
which is obvious nonsense.
Fourth, and finally,
insulators that are  close in chemical composition
and physical structure to specific high temperature superconductors,
have a measured finite relative permittivity greater than unity.
For example,
cuprate superconductors are insulators
if the doping fraction is less than 0.1 holes per CuO$_2$.\cite{CuO2}
It seems plausible that the small changes in composition
that turn these into conductors
do not much change  the residual static relative permittivity.
Instead the simpler interpretation is
that the infinite conduction permittivity
swamps any attempt to measure macroscopically
the finite residual static relative permittivity.

The most reliable way to measure the residual static dielectric permittivity
of a high temperature superconductor
may turn out to be by extrapolation
from an insulator of similar chemical and structural composition.

As mentioned in connection with Fig.~\ref{Fig:Tc},
a high monotonic-oscillatory transition temperature
relies upon the competing requirements
of high electron density and low relative permittivity.
One might speculate
that the reason for the prevalence of layered structures
amongst high temperature superconductors
is that they combine a high electron density
within the conducting planes
together with a low electron density and hence low polarisability
in the interlayer space.
At the level of dielectric continuua,
the consequent image charges increase the coupling
between the electrons in the dense layer\cite{Attard88}
and increase the temperature of the oscillatory transition
compared to a uniform dielectric medium.

Of course one has to be a little skeptical about the quantitative accuracy
of modeling a crystalline layered conductor as a homogeneous charge fluid.
On the one hand the universality of the monotonic-oscillatory transition
for the like-charge pair potential of mean force
at high coupling is robust and undeniable.
On the other hand, the quantitative prediction of the transition temperature
and the depth of the primary minimum cannot be taken too literally
for electrons in the layered crystalline solids
that are of interest in high temperature superconductivity.


This paper proposes that the oscillatory potential of mean force
that occurs at high coupling is responsible for Cooper pair formation
and superconductivity.
The BCS theory is predicated upon an attractive potential,
not an attractive potential of mean force.
The difference between the pair potential and the pair potential of mean force
is fundamentally the difference between quantum mechanics
and quantum statistical mechanics.
It is the latter, not the former, that is the appropriate theory
for condensed matter.
In an accompanying paper
I give a new quantum statistical mechanical theory for superconductivity
that shows explicitly how  Cooper pairs form and condense
depending upon the potential of mean force.\cite{Attard22c}
The theory uses the classical phase space
formulation of quantum statistical mechanics,\cite{Attard18,Attard21}
together with techniques recently developed for superfluidity.
\cite{Attard21,Attard22}




\end{document}